\documentclass[10pt,aps,prb,twocolumn, nofootinbib]{revtex4-1}

\usepackage{amssymb,amsmath}
\usepackage{subfig}
\usepackage{graphicx}
\usepackage{color}
\usepackage{hyperref}
\usepackage{listings}
\usepackage{float}
\usepackage{cleveref}
\usepackage{comment}
\usepackage[utf8]{inputenc}
\usepackage{algorithm}
\usepackage{amsmath}
\usepackage{mathtools}
\DeclarePairedDelimiter\abs{\lvert}{\rvert}
\usepackage{algpseudocode}
\hypersetup{
    bookmarks=true,         
    unicode=false,          
    pdftoolbar=true,        
    pdfmenubar=true,        
    pdffitwindow=false,     
    pdfstartview={FitH},    
    pdfauthor={Kevin Multani},     
    colorlinks=true,       
    linkcolor=blue,          
    citecolor=red,        
}
\graphicspath{{figures/}}

\algnewcommand\algorithmicforeach{\textbf{for each:}}
\algnewcommand\ForEach{\item[ \algorithmicforeach]}
\begin{document}

\definecolor{dkgreen}{rgb}{0,0.6,0}
\definecolor{gray}{rgb}{0.5,0.5,0.5}
\definecolor{mauve}{rgb}{0.58,0,0.82}
\def\andname{\hspace*{-0.5em},}

\lstset{frame=tb,
  	language=Matlab,
  	aboveskip=3mm,
  	belowskip=3mm,
  	showstringspaces=false,
  	columns=flexible,
  	basicstyle={\small\ttfamily},
  	numbers=none,
  	numberstyle=\tiny\color{gray},
 	keywordstyle=\color{blue},
	commentstyle=\color{dkgreen},
  	stringstyle=\color{mauve},
  	breaklines=true,
  	breakatwhitespace=true
  	tabsize=3
}

\title{Opinion dynamics and consensus formation in the Deffuant model with extremists and moderates in the population}

\author{L. Marconi$^1$, and
        F. Cecconi$^2$}
\affiliation{$^1$Advanced School in Artificial Intelligence, Brain, Mind and Society, \textit{science2mind} Association, Via San Martino della Battaglia 44, Rome, Italy}
\affiliation{$^2$LABSS - ISTC - CNR, Institute of Cognitive Sciences and Technologies\\ Via S. Martino della Battaglia 44, Rome, Italy}




\begin{abstract}
Among the different disciplines in the social, behavioral, and economic sciences, a fundamental class of problems is related to the prediction of the final state of the presence of individual opinions in a large population. The main aspects investigated in the opinion dynamics are related to the possibility of reaching an agreement and the time needed. In general, a consensus model allows to understand if a set of interacting agents can reach a consensus when choosing among several options: political vote, opinions, cultural features are well-known examples of them. Opinion dynamics models, then, can be seen as a sub-set of consensus models, when the options to choose from are opinions and the possible social consensus to be reached is the opinion agreement. Moreover, in large populations it is intuitive to find moderates, whose opinion is not at the extremes of the political opinion spectrum. Due to the wide range of their empirical presence, the study of their influence in the opinion dynamics processes is very important to enhance the comprehension of the political opinions evolution in large populations. The Deffuant model is aimed at studying consensus formation assuming that the opinion distribution of a population evolves via a sequence of random pairwise encounters. In this work we strive to extend the classical Deffuant model with the presence of moderates, in addition to the extremists, so as to study the emergence of consensus with different initial configurations of opinions and parameters. Therefore, we assess the role and the importance of the moderates for the appearance of consensus in the opinion dynamics processes within in large populations. We show that the population behavior is affected by the presence of a critical number of moderates, in specific configurations of the model parameters: when this threshold is reached the opinions in the population progressively converge.  

\end{abstract}

\maketitle

\section{\label{sec:one}Introduction}

In any social system, the interactions among the agents have a pivotal role in all the decision-making processes, influencing or stimulating opinion formation in populations, both of humans and animals. By everyday experience, one of the most important social phenomenon is the agreement: especially when it comes to a discussion involving different opinions, the process of reaching at least a partial agreement proved to be a basic catalyst of the evolution of the human thinking during time. Discussions involving divergent opinions or point of views allow the social actors to modify their perspective on the topics involved \citep{doi:10.1080/01621459.1974.10480137} \citep{doi:10.1086/228313}.\\

In the computational social sciences, plenty of models are aimed at studying opinion dynamics: often each model proposed can be modified with various variants. In general, a consensus model allows to understand if a set of interacting agents can reach a consensus when choosing among several options: political vote, opinions, cultural features are well-known examples of them. Opinion dynamics models, then, can be seen as a sub-set of consensus models, when the options to choose from are opinions and the possible social consensus to be reached is the opinion agreement. \\
The voter model \citep{10.2307/2959329}  \citep{GRANOVSKY199523}  \citep{Castellano_2003} \citep{PhysRevE.69.016109}  \citep{PhysRevLett.112.158701} describes a social stochastic process where each agent, represented by a node in a network, has to make a binary choice, randomly interacting with his neighbors and emulating them: the model is studied in order to understand, among other aspects, if the voters can eventually reach an agreement and how.\\
Beside the famous voter model, there are various other models studied in the opinion dynamics branch of the computational social science. The models can differ in the interaction mechanisms considered or in the features of the network structure of the agents involved. \\
While in the voter model the social mechanism examined is the imitation of peers, in the Sznajd model \citep{STAUFFER200293}, instead, the dynamics of the system is determined by two different rules: social validation and discord destroys. The former represents a partial agreement in a total neighborhood of agents when two (or more, depending on possible variants) adjacent nodes agree on one chosen opinion, while the latter represents the local division and arguing process when a block of adjacent neighbors disagree. As the methodological way of building up these models directly come from statistical mechanics, also some variants of the well-known Ising model can be used in order to study the social dynamics in response to phenomena like the social pressure. Other examples of consensus models were not born specifically in order to study opinion dynamics, but they (or their variants) are also used for this purpose. In the Axelrod's model  \citep{doi:10.1177/0022002797041002001} the stochastic process considered is the cultural dissemination: in this case, each node is characterized by a set of $F$ cultural features, each of which can assume $q$ states. Similarly to the voter model, the social influence is considered, but another mechanism studied in this model is homophily, which is relevant in the culture dynamics process and in the community formations. Other models, like Granovetter's \citep{doi:10.1086/226707}, investigate the possible presence of a critical threshold in the number of agreeing decision makers, which, if reached, determines the realization of a consensus.\\

The model introduced by Deffuant \textit{et al.} \citep{doi:10.1142/S0219525900000078} \citep{doi:10.1002/cplx.10031} can be considered as a bounded-confidence model, thus dealing with selective exposure, namely the agents' tendency to encourage information supporting their own viewpoints while 
omitting opposed arguments. The agents in the set, connected to
each other by an interaction network, own continuous opinions, with possible variations at each discrete time step according to the closeness to each other. In particular, the Deffuant model follow a sequential updating rule and can be seen as a discrete-time repeated game where the opinion distribution of the population evolves via a sequence of random pairwise encounters until the opinions convergence. The final state can either be composed by a single opinion or by multiple opinions. \\
When two agents meet, the possibility of a reciprocal inﬂuence is regulated by a parameter $\theta\geq0$, the tolerance of the individuals, according to the distance between their opinions. Moreover, another parameter $\mu\in\left(0,\frac{1}{2}\right]$ incorporates the willingness of an individual to consider a compromise with the opinion of another agent. This is sometimes called \textit{cautiousness parameter} and it tunes the speed of convergence.\\
The Deffuant model is specifically aimed at studying opinion-formation processes in large populations with small groups of agents reciprocally interacting, as pairwise interactions in any structured agents' network.\\

In spite of its appearing simplicity, the Deffuant model is
not analytically solvable in general. Monte Carlo simulations are mostly used to provide results about it. Numerical simulations for a few values of the cautiousness parameter show that consensus appears for large confidence bound values on complete graphs with probability close to $1$ in the large-population limit,
while multiple opinion groups persist at equilibrium
for low confidence bounds \citep{doi:10.1002/cplx.10031,doi:10.1002/cplx.20018,doi:10.1142/S0129183104006728}.\\
From an analytical perspective, some studies of the Deffuant
model used a density function that determines the agents’ density in opinion space \citep{LORENZ2005217,BENNAIM2003190}, but with restrictive assumptions.\\

In the opinion dynamics research area the role of moderates versus extremists, or at least of the extremists only, has been studied in a few works dealing with the Deffuant model or other models. Tsang and Larson \citep{10.5555/2615731.2615778} consider skeptical agents when included in a model taking in consideration the agents trust and empathy, leading them to be receptive toward other agents with similar opinions, so as to study the possible opinion convergence in different network configurations. Instead, Sobkowicz \citep{10.3389/fphy.2015.00017} study the occurrence of the extremist in a modified version of the Deffuant model, when the psychological structure of the agents' emotions is explicitly taken in consideration. Other studies try to study the Deffuant model exploring different formulations, like population-balance modeling \citep{article}.\\



The goal of this paper is to elucidate the contributions of the moderates for the appearance of consensus in the opinion dynamics processes within large populations, within some significant parameters configurations. Therefore, we extend the classical Deffuant model with the presence of moderates, in addition to the extremists, so as to study the emergence of consensus with different initial configurations of opinions and parameters.

Section~\ref{sec:one} delivers a basic mathematical introduction to the Deffuant model and provides insight into how the opinion dynamics mechanisms for the Deffuant model work. Moreover, we assess the role of the moderates in the agents' configurations.
Section~\ref{sec:two} describes the experimental setup for all the experiments conducted and assumptions made for each one. Section~\ref{sec:three} presents the results of the experimentation and provides the experimentally determined values for the critical number of moderates determining the appearance of consensus in large populations with different parameters and agents number. Section~\ref{sec:four} provides a conclusion with future considerations for further and more accurate experimentation.

\section{\label{sec:one}Theoretical Framework}
In the Deffuant model, at each discrete time step two neighboring
agents are randomly selected and interact in a pairwise manner. Only if their opinion difference is below a given threshold, the result of the interaction is a sort of compromise toward each other’s opinion. Otherwise, there is no modification in their opinions.
Considering a population of $N$ agents, the opinion space is $\left[a,b\right]\in R$. If agents $u$ and $v$ are selected uniformly at random and meet at time $t$, holding opinions $\left[a,b\right]\in R$ respectively, the update rule reads as follows:

\begin{equation}
\label{eq:one}
(\eta t(u),\eta t(v))=\begin{cases}
(a+\mu(b-a),b+\mu(a-b)) & \text{if}|a-b|\text{\ensuremath{\le}}\theta,\\
(a,b) & \text{otherwise}
\end{cases}
\end{equation} 
where $\eta t(u)$ denotes the opinion of agent $u$ at time $t$. 
Then, as two individuals approaching start interacting and discussing the topic in question, each one will only consider the opinion of the other agent as worth considering if it is close enough to their own personal belief. Closeness is measured by the parameter $\theta$. In this case, they will start a constructive debate and their opinions will symmetrically get closer to each other. In the special case $\mu=\frac{1}{2}$ the outcome of the interaction will be a complete agreement at the average of their previous opinions.\\
Therefore, the model allows an initial concentration of clusters of similar opinions and successively the final opinion states are reached by internal interactions. The general aim of the experiments on the model is to ﬁgure out for which values of the parameters $\theta$ and $\mu$ the agent set will result in one final opinion cluster (consensus) or split into several clusters (fragmentation).\\

In detail, the model exploits two characterizing parameters: the tolerance $\theta$ and the influence capacity $\mu$. The tolerance is built as a confidence bound describing a population’s resistance in front of diverse viewpoints. If the difference between the opinions of the two agents is lower than this threshold, their disagreement is reduced by making a compromise. Otherwise, the two agents maintain their current opinions after the interaction (if they actually  are willing to discuss the issue for real).\\ 



Instead, the parameter $\mu$ measures the influence capacity of the model, as a multiplier stating the relative agreement between the agents involved. It is also called a \textit{convergence parameter} \citep{doi:10.1142/S0219525900000078,doi:10.1002/cplx.10031,doi:10.1002/cplx.20018,doi:10.1142/S0129183105008126}, due to its role in specifying a population’s cautiousness in modifying their viewpoints.
For larger values of $\mu$ individuals are more
willing to make compromises. In the special case $\mu=0.5$,
each pair of interacting agents agree on the mean of their opinions,
whenever their opinion difference is below the confidence
bound. 

In the first formulation of the Deffuant model \citep{doi:10.1142/S0219525900000078}, it considered a ﬁnite number of agents having initial opinions uniformly distributed on $[0,1]$. Thus, a continuous opinion space is used, due to the possible and realistic variations of an individual’s viewpoint on a specific topic, e.g. politics.\\
Traditionally, opinion dynamics and opinion-formation processes have been studied considering discrete opinions,being a reasonable assumption in classical contexts, as in the voter model \citep{10.1093/biomet/60.3.581} \citep{10.2307/2959329}, where the vote decision is a binary choice.\\

Here we explicitly and extensively consider the so-called \textit{moderates}, namely the agents with non-extreme opinions. In our model, we have $N=N1+N2$ individuals, where $N1$ is the sum of all the extremists, with opinion either $0$ or $1$, and $N2$ is the number of all the moderates, whatever their opinions are. As an example, an agent is considered moderate if its opinion is $0.5$ or $0.125$, differently from an extremist with opinion $0$ or $1$.
Intuitively, if there are $N$ individuals, where $N-1$ have opinion $0$ and only one agent has opinion $1$, the probability to meet that agent is exactly $\frac{1}{N}$. Then, at each step, this probability can increase, according to the possible growth of the agents with opinion $1$ in the population.
In this work, we aim at assessing the role of the moderates in the model by considering different configurations of the parameters and of the number of agents in the set.\\
Therefore, let us consider four iconic situations, naming them with symbolic labels:
\begin{itemize}
    \item \textit{Centrality}: this is the case when $\mu=\theta=0.5$, namely both the tolerance to divergent opinions and the model's capacity to stimulate reciprocal influence are at the central point of their range.
    \item \textit{Debility}: here $\mu=\theta=0.25$, thus the model has both a low confidence bound and convergence parameter.
    \item \textit{Tolerability}: in this case $\theta=0.5$ and $\mu=0.25$, so the model shows medium tolerance and low influence capability.
    \item \textit{Susceptibility}: in this case $\mu=0.5$ and $\theta=0.25$ thus, conversely, the model shows medium influence capability and low tolerance.
\end{itemize}



Let us consider the first case (centrality), with the parameters be $\mu=\theta=0.5$ and two agents $p$ and $q$ be extremists, namely have the opinions either $\eta t(p)=\eta t(q)=0$, or, conversely, $\eta t(p)=\eta t(q)=1$. In both these two extreme cases any interaction between couple of such agents will not lead to any compromise: the same identical stance is maintained, due to the previous agreement. By extension, in general any meeting of couples of agents with identical opinions will not result in opinions evolution: e.g, this happens in the same party.\\
Now let us examine the encounter between other two agents $d$ and $s$, with $\eta t(d)=0$ and $\eta t(s)=1$. Even in this case, agents will not modify their previous viewpoint on a certain topic. Therefore, we can immediately observe that encounters among extremists are actually infertile, in the Deffuant model. It is then worth to notice that the presence of the moderates is already proven necessary to foster opinion dynamics in this model, at least when $\mu=\theta=0.5$. Then, in this case moderates trigger new opinions generation.\\

By extension, one could analytically compute the results in the other cases and explore the different situations occurring as the agents or the opinions vary in the population.\\
Unfortunately, the model is not analytically solvable, due to the  integro-differential equations describing it and the continuous opinion space. Thus, obtaining a closed-form solution is hard and we perform extensive experiments in the cases considered.\\

Initial opinions are assumed as independent and identically distributed according to the uniform distribution on the opinion space $[a,b]$. This assumption is well-known in literature, both in the original paper \citep{doi:10.1142/S0219525900000078} that introduced the Deffuant model, and in most later studies. This convention allows to deepen the comprehension of the classical model by providing numerical studies for results in the model’s variants. Nonuniform initial opinion distributions are considered, for example, in \citep{doi:10.1142/S0129183106010108}.\\

\section{\label{sec:two} Experimental Setup}

Extensive different experiments were conducted to determine the behavior of the individuals in terms of opinion evolution, in the four main parameters configurations considered. The experimental setup and execution were performed through NetLogo and the primary information collected over the course of the experiments were:
	\begin{itemize}\parskip0pt
        \item convergence parameter $\mu$
        \item tolerance $\theta$
        \item number of moderates 
        \item standard-deviation of the opinions 
        \item mean of the opinions 
        \item number of agents with opinion $1$ over the population
        \item agents with opinion $0$ over the population
        \item proportion of agents with opinions lower than $0.25$ 
        \item proportion of agents with opinions higher than $0.75$
    \end{itemize}

In preparation for all experiments, NetLogo was initialized by setting up the possibility to run the model many times as needed, systematically varying the model’s settings and then recording the results of each run. This process allows to iteratively modify the parameters after a fixed number of runs, exploring the different configurations and showing the emergent behaviors in the system.  Each time unit (\textit{tick}) corresponds to an execution of the imitation process: each agent randomly searches a partner and the Deffuant process starts. The total number of agents is $N = 500$, with half of the agents with opinion $0$ and half with opinion $1$. Thus, the initial configuration of the population is established with only extremists. Then, the algorithm allows the possibility to choose the number of moderates and to vary it in the different experiment, as described in the below Algorithm \ref{alg:Deffalg}.

\begin{algorithm}[H]
\caption{Deffuant model}
\label{alg:Deffalg}
\begin{algorithmic}[1]

\Procedure{Deffuant}{$N$}       
    \State System Initialization
    \State Create $N$ agents with opinion $o$
    \State Set $\mu$ and $\theta$
    \State Set $N/2$ agents with opinion $o = 0$
    \State Set $N/2$ agents with opinion $o = 1$
    \State Set $M \in N$ agents with opinion $o \in (0,1)$\\
    \ForEach {$(u,v) \in N$}\\
    \State Let $u$ and $v$ agents randomly meet
    \State Let ${o}_{u}$ the opinion of $u$
    \State Let ${o}_{v}$ the opinion of $v$\\
    \If{\begin{math}\abs{({o}_{u} - {o}_{v})} \leq \theta \end{math}}
        \State Let ${o}_{u} = {o}_{u} + \mu \times ({o}_{v} - {o}_{u}) $
        \State Let ${o}_{v} = {o}_{v} + \mu \times ({o}_{u} - {o}_{v})$
     \Else
        \State Do nothing
    \EndIf
\EndProcedure 

\end{algorithmic}
\end{algorithm}

The above algorithm allows to set the convergence parameter and the tolerance of the population, as well as the number $M$ of the moderates, and generates a simulation of the opinion dynamics \textit{à la Deffuant}.\\

We have performed $50$ runs for each configuration of the parameters $\mu$, $\theta$ and number of moderates.


The experiments conducted are summarized in Table~\ref{tab:experiments}. 

\begin{table}[h]
\caption{\label{tab:experiments}Maximum run and moderates number for each configuration of the parameters in the experiments.}
\scriptsize
\begin{ruledtabular}
\begin{tabular}{cll}
Minimum moderates number & Maximum moderates number & Parameters\\ \hline
50 & 350 & \begin{tabular}[c]{@{}l@{}}$\mu = 0.25$\\ $\theta = 0.25$\end{tabular}\\ \hline
50 & 350 & \begin{tabular}[c]{@{}l@{}}$\mu = 0.25$\\ $\theta = 0.5$\end{tabular}\\ \hline 
50 & 350 & \begin{tabular}[c]{@{}l@{}}$\mu = 0.5$\\ $\theta = 0.25$\end{tabular}\\ \hline
50 & 350 & \begin{tabular}[c]{@{}l@{}}$\mu = 0.5$\\ $\theta = 0.5$\end{tabular}\\ 
\end{tabular}
\end{ruledtabular}
\end{table}

\bigskip

The performed experiments were then designed based upon the theory discussed in Section~\ref{sec:one} in the different configurations of the parameters and of the number of the moderates in the population.

At the end of all the runs, we extracted four plots, able to represent the main aspects of the population behaviors, in the four parameters configurations examined:
\begin{itemize}
    \item mean of the standard deviations of the agents' opinions  in the different runs for each  of the parameters configurations 
    \item mean of the standard deviation of the agents' opinions in the third quartile  in  the  different  runs  for  each  of the parameters configurations
    \item mean of the proportions of agents with opinion $1$ in the total amount of the population, when varying the moderates number
    \item mean of the proportions of agents with opinion $0$ in the total amount of the population, when varying the moderates number
    \item mean of the proportions of the agents with opinions higher than $0.75$ in the total amount of the population, when varying the moderates number
    \item mean of the proportions of the agents with opinions lower than $0.25$ in the total amount of the population, when varying the moderates number
\end{itemize}

\section{\label{sec:three}Results and Discussion}
\setcounter{subsubsection}{0}
\renewcommand*{\theHsubsubsection}{chX.\the\value{subsubsection}} 

The results collected are represented in the following figures.\\

\begin{figure}[H]
\centering
\includegraphics[width=1.0\linewidth]{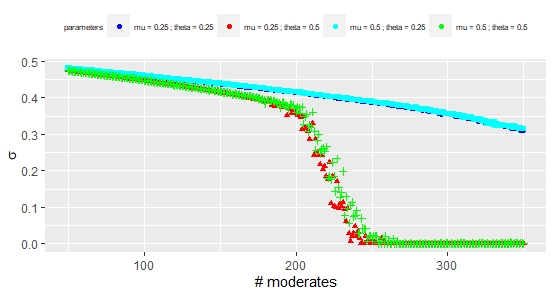}
\caption{Results of the calculation of the mean of the standard deviations of the opinions in the different runs for each of the parameters configurations}
\label{fig:mean_of_sd}
\end{figure}

Here we can see that the mean of the standard deviations of the  agents’ opinions in the different runs for each of the parameters configurations definitively shows a phase transition occurring in the agents behavior in the opinion space, leading to opinions convergence. This happens in the situations of \textit{tolerability} and \textit{susceptibility}, namely when $\mu = 0.5$. The order parameter of the phase transition is the mean of the standard deviations of the opinions. The opinions behavior shows the phase transition when moderates reach the critical value of $200$. Therefore, after the critical value of moderates in the populations, the difference among opinions is progressively reduced to zero.\\

The behavior is clearly different relating to two macro-cases:
\begin{itemize}
    \item \textit{centrality} and \textit{tolerability}: here $\theta=0.5$, namely the tolerance is enough to trigger the phase transition;
    \item\textit{debility} and \textit{susceptibility}: here $\theta=0.25$ and the phase transition does not appear due to the lower tolerance.
\end{itemize}

Therefore, it is clear that there is a critical threshold in the moderates number, determining the approach to consensus after it within specific parameters conditions. When the critical number of moderate individuals is reached in the population, the consensus formation is fostered.\\

Then, when the number of the moderates increases in the population, the variation of the opinions monothonically decreases, until it reaches a critical value and the decrease falls rapidly.\\

This behavior implies that, when the number of the moderates increases, given some specific conditions on the values of $\mu$ and $\theta$, the whole population converges towards a specific opinion. Thus, for specific configurations of the model parameters, there is a threshold determining the point when opinion convergence starts.\\

As a consequence, it is worth to notice that, in particular conditions of $\mu$ and $\theta$, the other moderates or extremists progressively disappear in the population, after the critical threshold is reached.\\

In the following figures \ref{fig:agents_op_1} and \ref{fig:agents_op_0} we collect the results for the mean of the proportions of agents with opinion $1$ in the total amount of the population, varying the moderates number. 

\begin{figure}[H]
\centering
\includegraphics[width=1.0\linewidth]{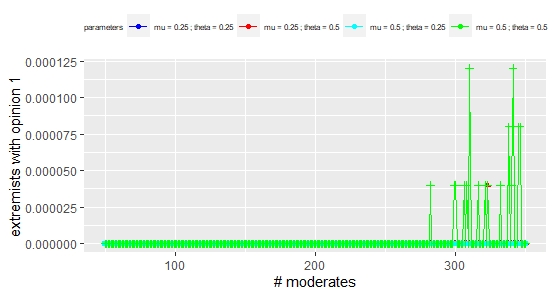}
\caption{Results of the calculation of the mean of the proportions of agents with opinion $1$ in the total amount of the population,  varying the moderates number}
\label{fig:agents_op_1}
\end{figure}
Thus, extremists are not present in the population when varying the moderates number, except for some sporadic agents, again when $\theta = 0.5$.\\

The situation is completely symmetrical as regards the mean of the proportions of agents with opinion $0$ in the total amount of the population, always when varying the moderates number, represented in the following figure \ref{fig:agents_op_0}.

\begin{figure}[H]
\centering
\includegraphics[width=1.0\linewidth]{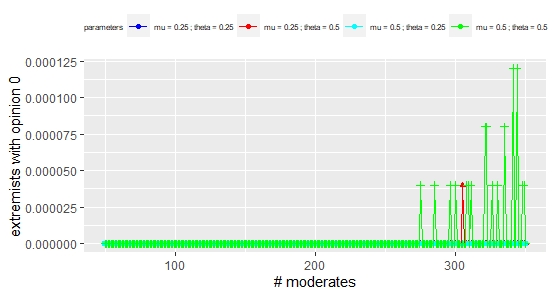}
\caption{Results of the calculation of the mean of the proportions of agents with opinion $0$ in the total amount of the population,  varying the moderates number}
\label{fig:agents_op_0}
\end{figure}

Thus we notice that the number of extremists in the population is progressively decreased, up to the end of the different runs of the experiments.\\

Nevertheless, again some sporadic extremist appears at the end, when $\theta = 0.5$.

Instead, the following figures \ref{fig:agents_quart} and \ref{fig:agents_quart_0.25} represent the results for the mean of the proportions of the agents with opinions higher than $0.75$ and lower than $0.25$ respectively in the total amount of the population, when varying the moderates number. 

\begin{figure}[H]
\centering
\includegraphics[width=1.0\linewidth]{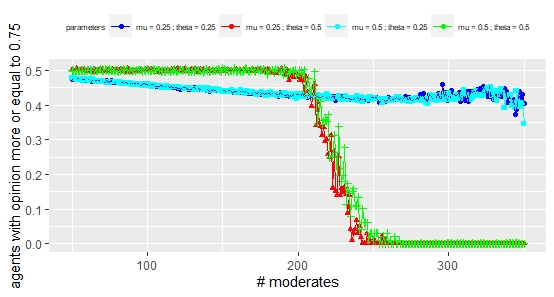}
\caption{Results of the calculation of the mean of the proportions of the agents with opinions higher than $0.75$ in the total amount of the population, when varying the moderates number}
\label{fig:agents_quart}
\end{figure}

\begin{figure}[H]
\centering
\includegraphics[width=1.0\linewidth]{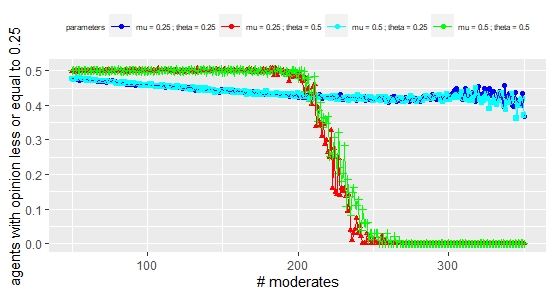}
\caption{mean of the proportions of the agents with opinions lower than $0.25$ in the total amount of the population, when varying the moderates number}
\label{fig:agents_quart_0.25}
\end{figure}

Therefore, when the number of moderates increases in the population the opinions tend to concentrate and a consensus is likely to appear. Indeed, the agents' behavior shows a second-order phase transition  occurring when the critical number of $200$ moderates is reached, leading to opinions convergence. Then, in our system the order parameter is the standard deviations of the opinions in the agents population.\\

Nevertheless, there is a non-zero probability to encounter some very sporadic extremists at the end of the simulation. Specifically, we see that in the cases where $\theta = 0.5$ (\textit{centrality} and \textit{tolerability}) and there are enough moderates in the population, we notice the presence of some extremists with opinion $0$ or $1$. This implies that these agents never interacted with moderates, buy only with other extremists with the same opinion.\\

\section{\label{sec:four}Conclusion}
In this work we have studied and extended the classical Deffuant model including explicitly moderates agents in large sample populations, in addition to the extremists, with the aim to analyse the emergence of consensus and opinions convergence within some significant initial configurations of opinions and parameters. Therefore, we have assessed the role and the importance of the moderates for fostering consensus in opinion dynamics processes within in large populations. We have shown that the population behavior is affected by the presence of a critical number of moderates, in specific parameters configurations: when this threshold is reached the opinions in the population progressively converges. This happens when the tolerance in the model is sufficiently high to foster the phase transition.\\

Paradoxically, we have discovered that, in the same parameters conditions, some extremists are still present at the end of the experiments. This emerging situation implies the possibility that, even in a highly-moderate environment, some clusters of non-influenceable extremists tend to remain.\\

This behavior is certainly counter-intuitive and poses socio-political implications in the treatment of extremism. Indeed, extremism is a challenges for democratic societies and countries. Nevertheless, this study focuses the attention on the possible contribution of the moderates themselves to the appearance of extremism: this requires attention in social systems when extremists and moderates interact, e.g. social networks (which play a significant role in real opinion formation and sharing processes).\\ Of course, these results are not always present in dynamics, but only in significant and specific conditions on the population and opinions structure and parameters. However, the results obtained require attention and more study to increase awareness on the opinion dynamics and the respective behaviors of extremists and moderates in large populations. In particular, further investigations are needed to study whether and how these resistant clusters of highly-extremists remain in the population, as well as their influence towards possible expansions.\\

Finally, the phase transition is certainly relevant for better understanding the role of the moderates in the opinion dynamics for large populations. Nevertheless, deeper study is required to better examine its presence and characteristics in more environmental conditions.\\

Improvements can be made for future experimentation. Possible improvements would include extend the study with other significant configurations of parameters to be found. Other possible future steps include further modifications of the model formulation itself, designing other parameters for depicting behaviors like how much agents tend to be influenced from media or social media. Moreover, a relevant empowerment of the model would integrate formulations from other works or opinion dynamics model, to strengthen the reality representation. Through a more sophisticated modeling process simulation results would be enhanced, due to the possibility to better model real situations. Additionally, carrying out the experiment in a more controlled environment would help tune noise, opinions fluctuations and other possible non-linear effects in the simulations. 
If one can control as many of these as possible, a characteristic representation for the emergent behaviors in terms of opinions evolution can be better determined. Moreover, different initial conditions and parameters representations could help model real-life situations of consensus formations.

\bibliography{bib}
\end{document}